\documentstyle[fleqn,espcrc2_my,epsf]{article}

\newcommand{\AmS}{{\protect\the\textfont2
  A\kern-.1667em\lower.5ex\hbox{M}\kern-.125emS}}

\title{Heavy quark masses from the $Q\bar Q$ threshold and the 
       upsilon expansion}

\author{A. H. Hoang\address{
   Theory Division, CERN, CH-1211 Geneva 23, Switzerland}%
        }

\begin{document}


\thispagestyle{empty}
\begin{minipage}{16cm}
\begin{center}

\begin{flushright}
{\bf CERN-TH/99-272}\\
{\bf hep-ph/9909356}\\
{\bf September 1999}\\
\end{flushright}
\vspace{1.0cm}
\begin{center}
  \begin{Large}\bf
Heavy Quark Masses from the $Q\bar Q$ Threshold\\[2mm] and the 
       Upsilon Expansion
  \end{Large}
  \vspace{1.2cm}

\begin{large}
 A.~H.~Hoang
\end{large}

\vspace{.5cm}
\begin{it}
Theory Division, CERN,\\
   CH-1211 Geneva 23, Switzerland\\[.5cm]
\end{it}

  \vspace{1.5cm}
  {\bf Abstract}\\
\vspace{0.3cm}

\noindent
\begin{minipage}{14.0cm}
\begin{small}
Recent results from studies using half the perturbative mass of heavy 
quark-antiquark $n=1$, ${}^3S_1$ quarkonium as a new heavy quark mass
definition for problems where the characteristic scale is smaller than
or of the same order as the heavy quark mass are reviewed. In this new
scheme, called the $1S$ mass scheme, the heavy quark mass can be
determined very accurately, and many observables like inclusive B
decays show nicely converging perturbative expansions. Updates on
results using the $1S$ scheme due to new higher order calculations are
presented.   
\end{small}
\end{minipage}
\end{center}
\setcounter{footnote}{0}
\renewcommand{\thefootnote}{\arabic{footnote}}
\vspace{1.7cm}
\begin{it}
Talk given at High Energy Physics International \\
       Euroconference on Quantum Chromo Dynamics - QCD '99, \\
       Montpellier, France, 7-13 July 1999. 
\end{it}
\vspace{3.2cm}
\begin{flushleft}
{\bf CERN-TH/99-272}\\
{\bf September 1999}
\end{flushleft}
\end{center}
\end{minipage}

\newpage
\pagestyle{plain}
\setcounter{page}{1}


\begin{abstract}
Recent results from studies using half the perturbative mass of heavy 
quark-antiquark $n=1$, ${}^3S_1$ quarkonium as a new heavy quark mass
definition for problems where the characteristic scale is smaller than
or of the same order as the heavy quark mass are reviewed. In this new
scheme, called the $1S$ mass scheme, the heavy quark mass can be
determined very accurately, and many observables like inclusive B
decays show nicely converging perturbative expansions. Updates on
results using the $1S$ scheme due to new higher order calculations are
presented.   
\end{abstract}

\maketitle

\section{INTRODUCTION}

The top and bottom quark masses are very important phenomenological
quantities. Two prominent examples which illustrate that we need to
know them to a high degree of precision are virtual top quark effects
in electroweak precision observables and B decay phenomenology: the
top quark indirectly affects the relation between the $W$, $Z$ masses,
and the weak mixing angle $\theta_W$ through loop
effects which are usually parameterised by the quantity $\Delta
\rho\sim m_t^2 G_F$. Future improvements in the determination of $M_W$ at
the Large Hadron Collider (LHC) and Linear Collider (LC) make it
desirable to push the error in the top quark 
mass much below the level of a GeV in order to get stringent bounds on
the Higgs boson mass which enters the relation among the electroweak
precision observables only logarithmically. Thus
the analysis of electroweak precision observables is complementary to
direct Higgs searches and provides an important test of electroweak
symmetry breaking. The bottom quark mass enters the inclusive B meson
decay rates as the fifth power, $\Gamma\sim m_b^5 |V_{CKM}|^2$. Thus,
the errors in the bottom quark mass should be at the percent level
(i.e. not more than 50 MeV), if CKM matrix elements like $V_{cb}$
shall be determined with an error of a few percent from inclusive
decays.  

Using continuum QCD and perturbative methods the most accurate and
precise determinations of the top and bottom quark masses have and
will come from observables involving the $t\bar t$ and $b\bar b$
thresholds. Whereas hadron colliders, which determine the top quark
mass from a reconstructed b-W invariant mass distribution, will have
a very hard time to reduce the top mass error below 2 GeV due to large
systematic uncertainties, a lineshape scan of the total (colour
singlet) $t\bar t$ 
cross section close to threshold at the LC will easily determine the
top mass with a combined statistical and systematical {\it experimental}
uncertainty of order 100 MeV~\cite{Peralta1}.
The question is whether one can provide a theoretical description of
the threshold lineshape which allows for {\it theoretical}
uncertainties in the top mass extraction of the same order (or maybe
better). For the bottom quark
mass, on the other hand, the most precise determinations come from sum
rule calculations using the experimental data on the $\Upsilon$
mesons~\cite{PDG}. A quick look at the presently available bottom
quark mass determinations~\cite{PDG}, however, seems to indicate that
a bottom quark mass uncertainty of around 50 MeV is out of
question. Observing the spread of numbers given in~\cite{PDG} an
uncertainty of 150-200 MeV seems to be more realistic. 

On the other hand,
when talking about quark masses we have to keep in mind 
that, due to confinement, they are not observables, but parameters
multiplying the bilinear $\bar \psi\psi$ operators in the QCD
Lagrangian. Thus, they are always determined indirectly, and our
ability to determine them with high precision and their usefulness
for practical applications can depend on the cleverness of their
definition.
In this talk I report on recent studies using half the perturbative
contributions of a heavy quark-antiquark $n=1$, ${}^3S_1$ bound state
as a new heavy quark mass definition. This new scheme is called the
{\it $1S$ scheme}~\cite{Hoang2,Hoang4}. 
The $1S$ mass is a {\it short-distance mass}, i.e. it
does not contain an ambiguity of order $\Lambda_{QCD}$ and the problem
of large higher order corrections associated with a pole in the Borel
transform at $u=1/2$ like the pole mass. But, unlike the well known
$\overline{\mbox{MS}}$ mass, which we might consider as the proto-type
of a short-distance mass, the $1S$ mass is specialised for problems
where the characteristic scale is smaller than the quark mass -- a
region where the $\overline{\mbox{MS}}$ mass loses its conceptual
meaning. By construction, the $1S$ scheme is the optimal choice for
problems involving non-relativistic $t\bar t$ and $b\bar b$ systems,
and one can expect that the $1S$ mass can be determined from them with
small uncertainties. I will demonstrate the advantages of the $1S$
mass compared to the pole and the $\overline{\mbox{MS}}$ scheme for
the NNLO calculations of the total $t\bar t$ cross section close to
threshold at the LC~\cite{Hoang2} and a sum rule determination of the
bottom quark mass~\cite{Hoang4}.  
However, the $1S$ mass scheme also works
well for non-$Q\bar Q$ problems like inclusive $B$ meson decays. It
also allows for a more refined determination of the
$\overline{\mbox{MS}}$ mass. 
If the $1S$ scheme would not be applicable for non-$Q\bar Q$ system it
would be of little practical value. In
order to apply the $1S$ mass scheme to non-$Q\bar Q$ systems a
modified perturbative expansion, called the 
{\it upsilon expansion}~\cite{Hoang5}, has to be employed. 
I hope that the $1S$ scheme can contribute to the general
acceptance that the desired top and bottom quark mass uncertainties
mentioned above are realistic and can indeed be achieved, although the
$1S$ scheme is certainly not the only way to achieve this aim. At the 
end of this talk I will also comment on other low scale short-distance
masses that can be found in literature and their relation to the $1S$
mass.

\section{THE $1S$ MASS}

The $1S$ heavy quark mass is defined as half the
perturbative contribution of a  $J^{PC} = 1^{--}$,
${}^3\!S_1$ $Q\bar Q$ ground state mass. Expressed in terms of the 
pole mass the $1S$ mass at NNLO in the non-relativistic
expansion reads
($a_s=\alpha_s^{(n_l)}(\mu)$),~\cite{Pineda1,Melnikov1}  
\begin{equation}
M^{1S} = M^{pole}[
1 -
\epsilon\,\Delta^{\mbox{\tiny LO}}
-\epsilon^2\,\Delta^{\mbox{\tiny LO}}\delta^1
-\epsilon^3\,\Delta^{\mbox{\tiny LO}}\delta^2
]
\,,
\label{M1Sdef}
\end{equation}
where
\begin{eqnarray}
\Delta^{\mbox{\tiny LO}} & = &
\mbox{$ \frac{C_F^2 a_s^2}{8}
$}
\,,
\label{deltaLO}
\\
\delta^1 & = & 
\mbox{$
(\frac{a_s}{\pi})[
\beta_0( L + 1 ) + \frac{a_1}{2} ]
$}
\,,
\label{deltaNLO}
\\ 
\delta^2 & = &
\mbox{$
(\frac{a_s}{\pi})^2 [
\beta_0^2 ( \frac{3}{4} L^2 +  L + 
         \frac{\zeta_3}{2} + \frac{\pi^2}{24} +
         \frac{1}{4} ) 
$}
\nonumber
\\ & &
\mbox{$ \quad
+ \beta_0 \frac{a_1}{2} ( \frac{3}{2} L + 1 )+
\frac{\beta_1}{4} ( L + 1 )
$} 
\nonumber
\\ & &
\mbox{$ \quad
+ \frac{a_1^2}{16} + \frac{a_2}{8} + 
( C_A - \frac{C_F}{48} ) C_F \pi^2 ]
$}
\,,
\label{deltaNNLO}
\\
L & \equiv & 
\mbox{$
\ln (\frac{\mu}{C_F a_s M^{pole}})
$}
\,,
\end{eqnarray}
and 
\begin{eqnarray}
a_1 & = &
\mbox{$
\frac{31}{9} C_A - \frac{20}{9} T n_l
$}
\,,
\nonumber
\\
a_2 & = & 
\mbox{$
( \frac{4343}{162}+4 \pi^2-\frac{\pi^4}{4}
 +\frac{22}{3} \zeta_3 ) C_A^2 
$}
\nonumber
\\ & &
\mbox{$ \quad
- ( \frac{1798}{81}+\frac{56}{3} \zeta_3 ) C_A T n_l
$}
\nonumber
\\ & &
\mbox{$ \quad
- ( \frac{55}{3}-16 \zeta_3 ) C_F T n_l 
+ ( \frac{20}{9} T n_l )^2
$}
\,.
\end{eqnarray}
The constants $\beta_0=11-\frac{2}{3}n_l$ and 
$\beta_1=102-\frac{38}{3}n_l$ are the one- and two-loop
coefficients of the QCD beta function and the constants
$a_1$~\cite{Fischler1,Billoire1} and $a_2$~\cite{Schroeder1}
the non-logarithmic one- and two-loop corrections to the static
colour-singlet heavy quark potential in the pole mass scheme,
$V^{Coul}$.
All $n_l$ light quarks are treated as massless.
In Eq.~(\ref{M1Sdef}) we have labelled the contributions at LO, NLO and
NNLO in the non-relativistic expansion by powers $\epsilon$,
$\epsilon^2$ and $\epsilon^3$, respectively, of the auxiliary
parameter $\epsilon=1$. The meaning will become clear when we
introduce the upsilon expansion later in this talk.

$M^{1S}$ is a short-distance mass because it contains, by
construction, half of the total static energy $\langle 2
M_b^{pole} + V^{Coul}\rangle$ which can be proven to be free of
ambiguities of order $\Lambda_{QCD}$~\cite{Hoang0,Beneke0} (see
also~\cite{Uraltsev1}). The fact that the static potential is
sensitive to scales below the inverse Bohr
radius~\cite{Appelquist1} does not lead to ambiguities because
the $1S$ mass also contains the physical perturbative contributions
from momenta below the inverse Bohr radius.

\section{APPLICATION TO $Q\bar Q$ SYSTEMS -- 
$1S$ MASS DETERMINATION}

Within the last two years there has been significant progress in our
ability to calculate higher order corrections to non-relativistic
$Q\bar Q$ systems. For the case that the average energy of the quarks
is (much) larger than $\Lambda_{QCD}$ NNLO corrections
(i.e. corrections of order $\alpha_s^2$, $\alpha_s v$ and $v^2$) are
now available for the production cross section of $Q\bar Q$ pairs in
the threshold region. The conceptual framework in which those
perturbative calculations can be organised in an economical way is
(P)NRQCD, an non-relativistic effective field theory of QCD. Two talks
about this subject are given on this conference~\cite{NRQCDtalks}. The
newly available NNLO corrections clearly demonstrate the need for the 
introduction of a {\it low scale short-distance mass} like the $1S$
mass, if the desired quark mass uncertainties mentioned before shall
be achieved. 

\subsection{$t\bar t$ production at threshold}

\begin{figure}[t!] 
\begin{center}
\leavevmode
\epsfxsize=3cm
\leavevmode
\epsffile[220 580 420 710]{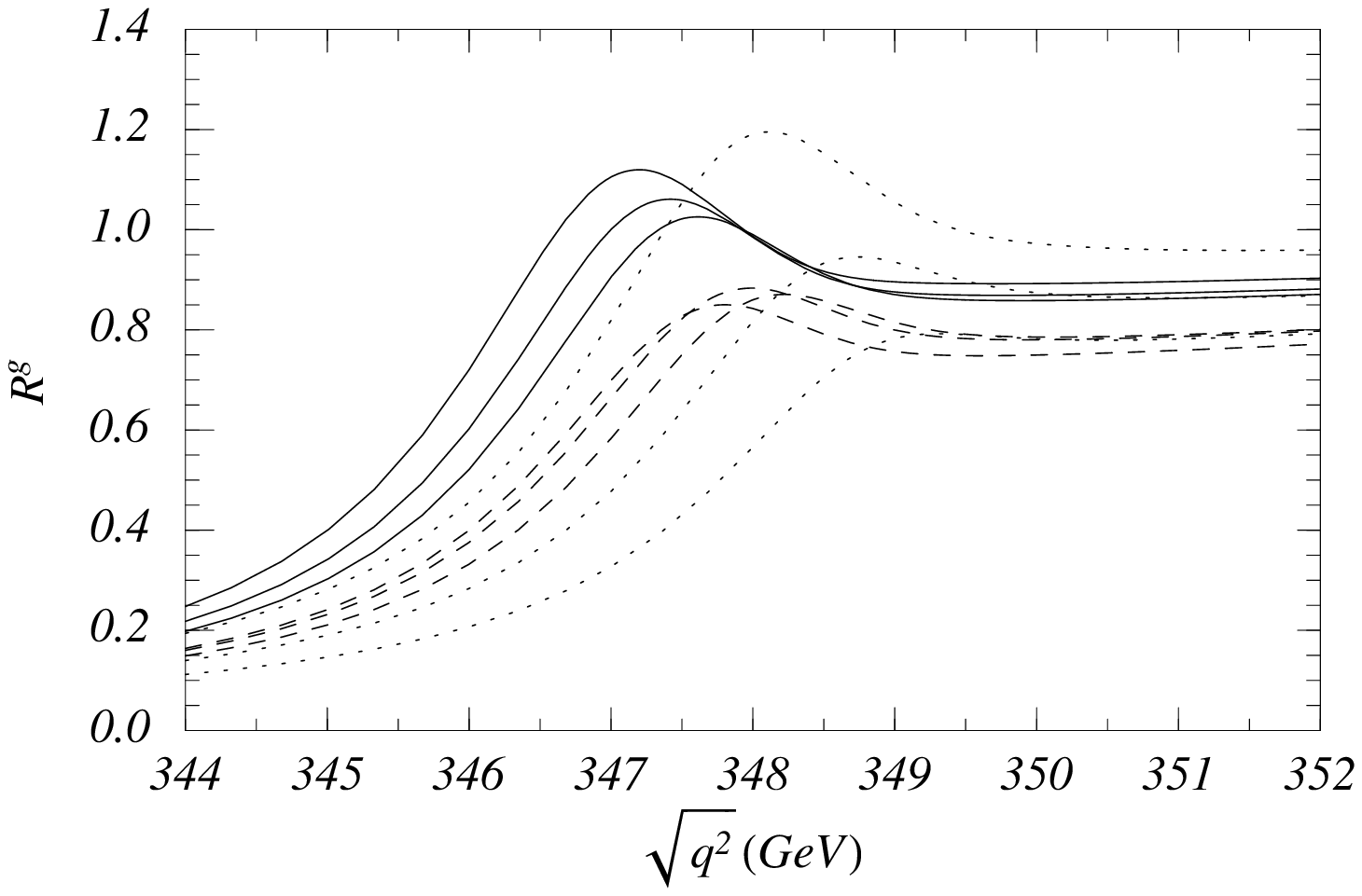}
\vskip 2.1cm
\epsfxsize=3cm
\leavevmode
\epsffile[220 580 420 710]{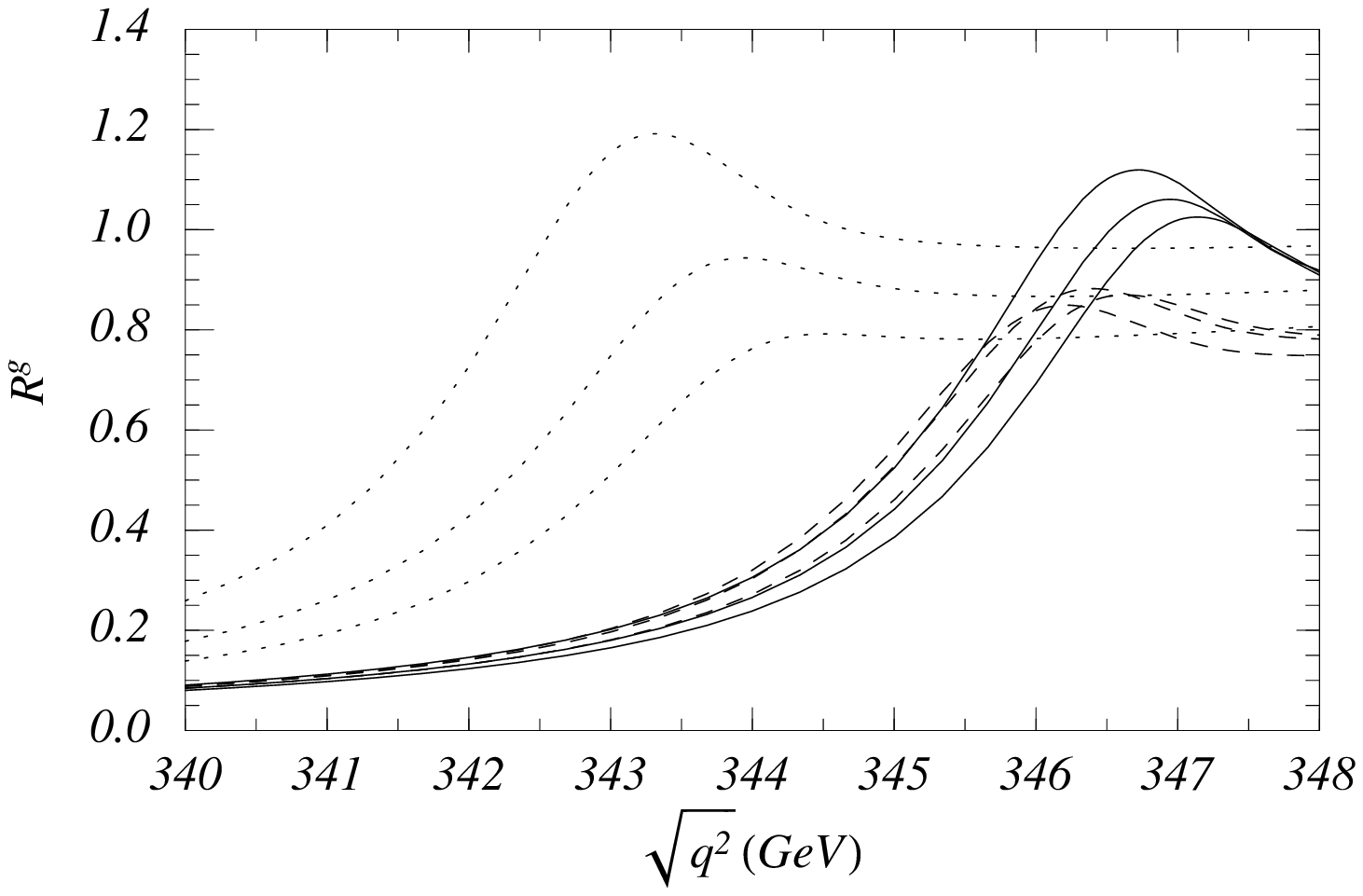}
\vskip 2.2cm
\epsfxsize=3cm
\leavevmode
\epsffile[220 580 420 710]{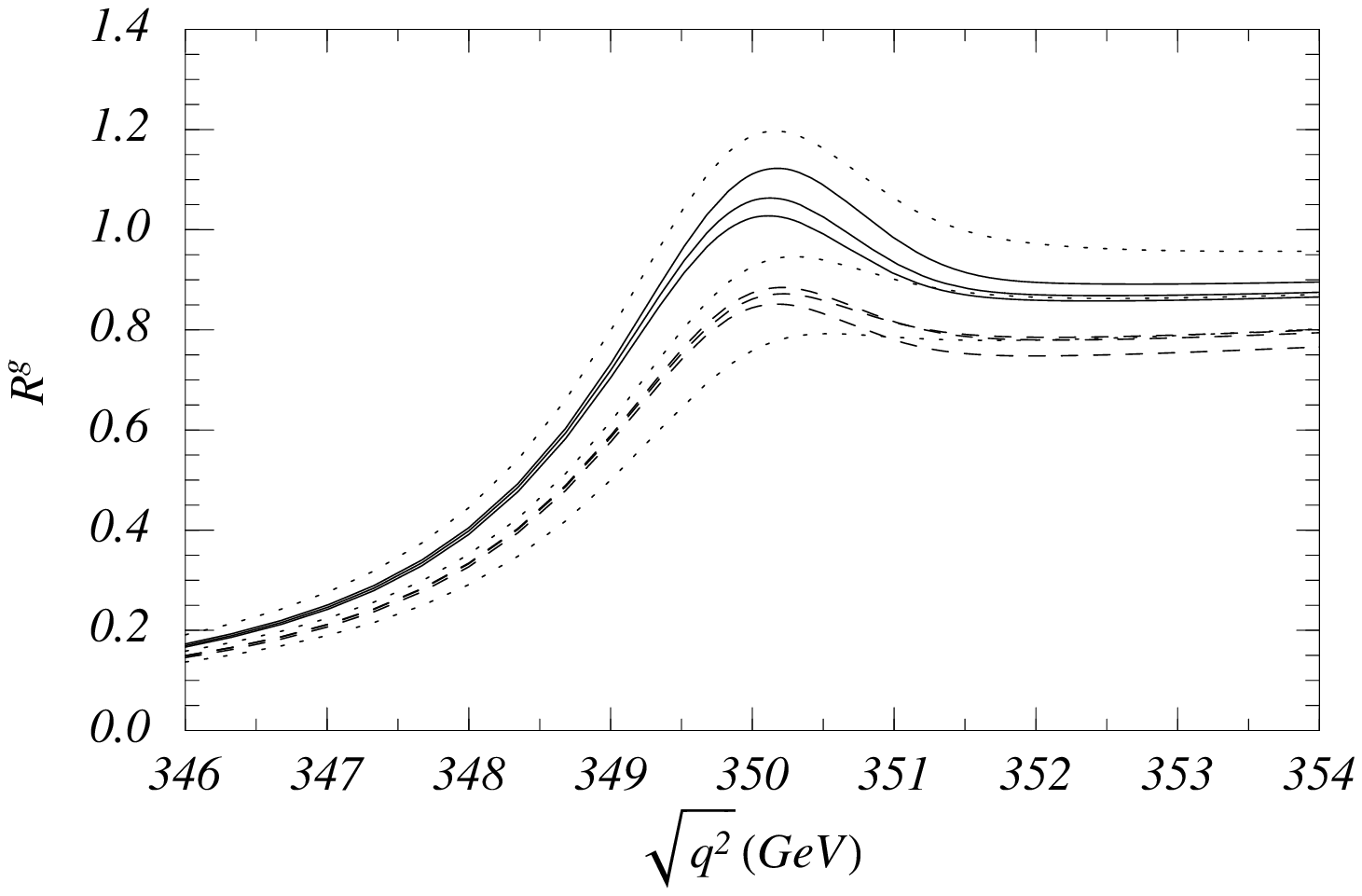}
%
%
\vskip  1.cm
 \caption{\label{figttbar}
The total normalised vector-current-induced $t\bar t$ cross section at
the LC versus the c.m. energy in the threshold regime at LO (dotted
curves), NLO (dashed) and NNLO (solid) in the pole (upper figure),
$\overline{\mbox{MS}}$ (middle) and $1S$ (lower) mass schemes for  
$\alpha_s(M_Z)=0.118$ and $\mu=15$, $30$, $60$~GeV. The plots have
been generated from results obtained in~\cite{Hoang1,Hoang2}. NNLO
calculations for the cross section have also been carried out
in~\cite{Melnikov2,Yakovlev1,Beneke1,Sumino1,Penin1}. 
}
 \end{center}
\end{figure}
In Figs.~\ref{figttbar} the total photon-mediated $t\bar t$
cross section in the threshold region at the LC normalised to the muon
pair cross section,
\begin{equation}
\mbox{$
R^\gamma\,\equiv\,
\frac{\sigma(e^+e^-\to\gamma^*\to t\bar t)}
     {\sigma(e^+e^-\to\gamma^*\to \mu^+\mu^-)}
\,,
$}
\end{equation}
 is displayed at LO, NLO and NNLO for three
different renormalisation scales. The three figures show the cross
section in the pole (upper figure), the
$\overline{\mbox{MS}}$ (middle) and the $1S$ (lower) mass schemes for
$(M_t^{pole},\overline 
m_t(\overline m_t),M_t^{1S})=(175,165,175)$~GeV. To implement the
$\overline{\mbox{MS}}$ mass scheme (here and also in the $\Upsilon$
sum rule analysis) the upsilon expansion
(Sec.~\ref{subsectionupsilon}) has been 
employed. We see that in the pole mass scheme the location of the
peak\footnote{
The $t\bar t$ cross section does not have resonances in the threshold
regime because the large top width $\Gamma_t\approx 1.5$~GeV
smears them out. Only the $1S$ resonance remains visible as
a slight enhancement of the cross section.
}
does not show any sign of convergence. In the $\overline{\mbox{MS}}$
mass scheme the peak location converges, but at the cost of even larger
corrections. The figures show that it is very hard to determine the
pole or the $\overline{\mbox{MS}}$ mass with theoretical uncertainties
below half a GeV. Compared to the best results expected from hadron
colliders the situation is not bad, but we can do much better. 
The situation improves dramatically in the $1S$
scheme, where the peak position is absolutely stable. Realistic
simulation studies~\cite{Peralta1} have shown that the $1S$ mass can be
extracted from the threshold scan with theoretical uncertainties of
order 100 MeV. Those studies have taken into account beamstrahlung
effects that lead to a smearing of the cross section and the fact that
the remaining normalisation uncertainties affect the top mass
determination.

\subsection{$\Upsilon$ mesons}

The $\Upsilon$ sum rules relate moments of the the correlator of two
electromagnetic bottom quark currents 
\begin{equation}
P_n^{th} \, = \,
\mbox{$
\frac{4\,\pi^2}{9\,n!\,q^2}
( \frac{d}{d q^2} )^n\,{\Pi^{b\bar b}}_\mu^{\,\,\mu}(q)
\big|_{q^2=0}
$}
\end{equation}
to a dispersion integral over the total $b\bar b$ production cross
section in $e^+e^-$ annihilation, 
\begin{equation}
P_n \, = \,
\mbox{$
\int^\infty_{\sqrt{s}_{min}} \frac{d s}{s^{n+1}}\,R^{b\bar b}(s)
$}
\,.
\end{equation}
For values of $n$ between about 4 and 10 the moments are saturated by
the non-relativistic $b\bar b$ bound states and, at the same time, can
be calculated perturbatively at NNLO in the non-relativistic
expansion. Choosing $\alpha_s$ as an input, and assuming global
duality, the $b$ quark mass can be extracted from fits to the
experimental data~\cite{Hoang3,Hoang4}. 
\begin{figure}[t!] 
\begin{center}
\leavevmode
\epsfxsize=3cm
\epsffile[220 580 420 710]{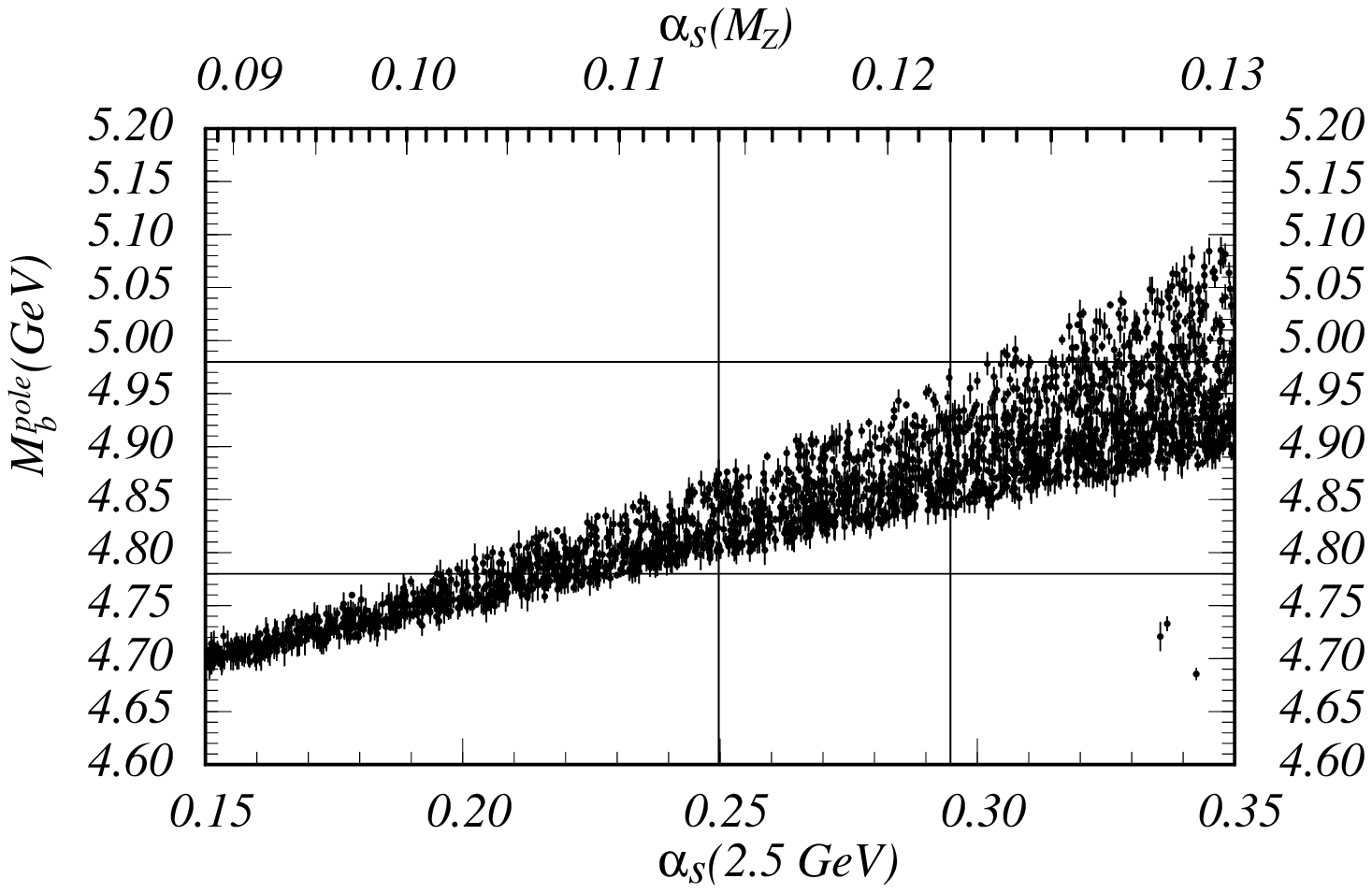}
\vskip 2.3cm
\epsfxsize=3cm
\leavevmode
\epsffile[220 580 420 710]{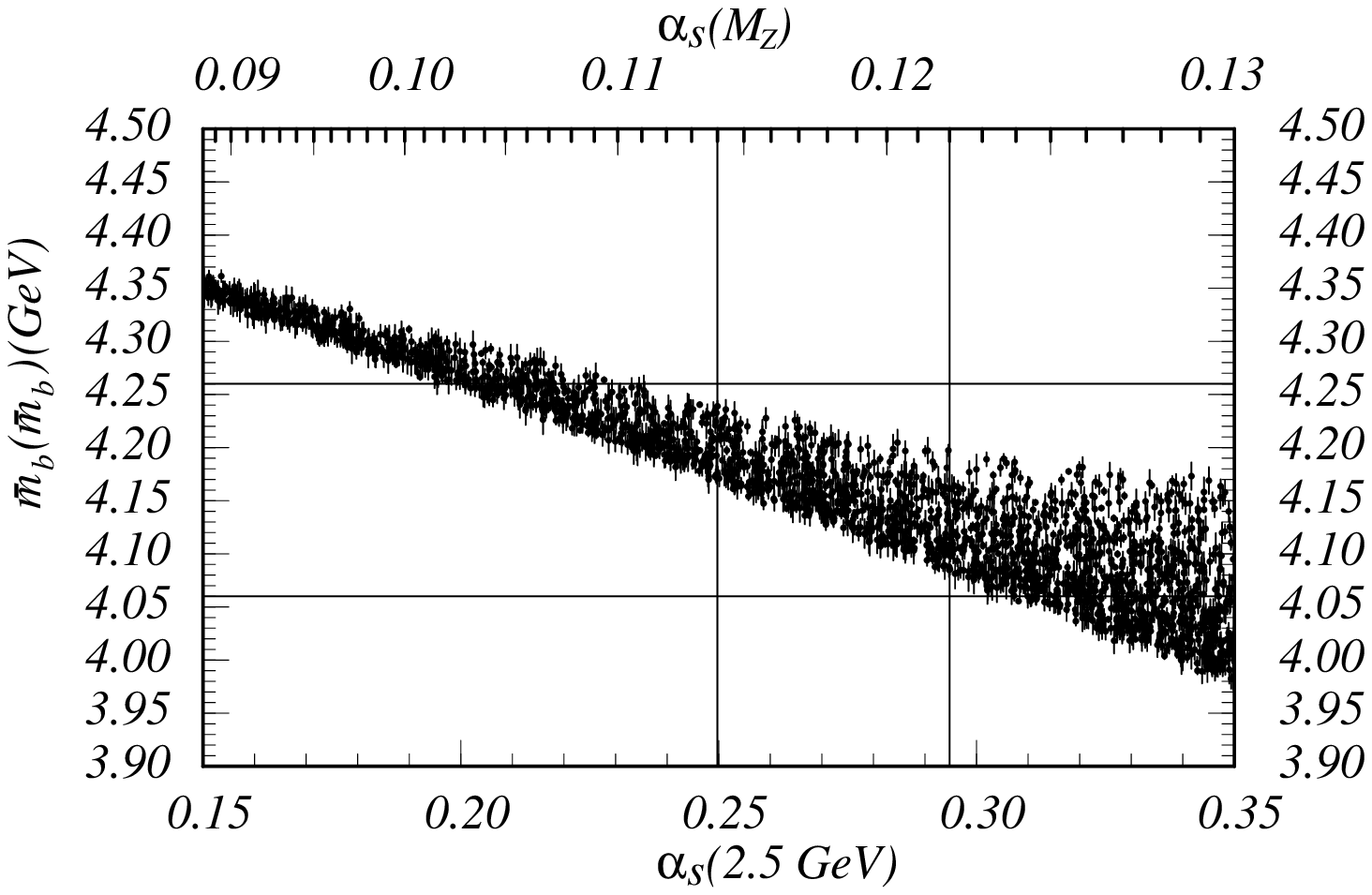}
\vskip 2.3cm
\epsfxsize=3cm
\leavevmode
\epsffile[220 580 420 710]{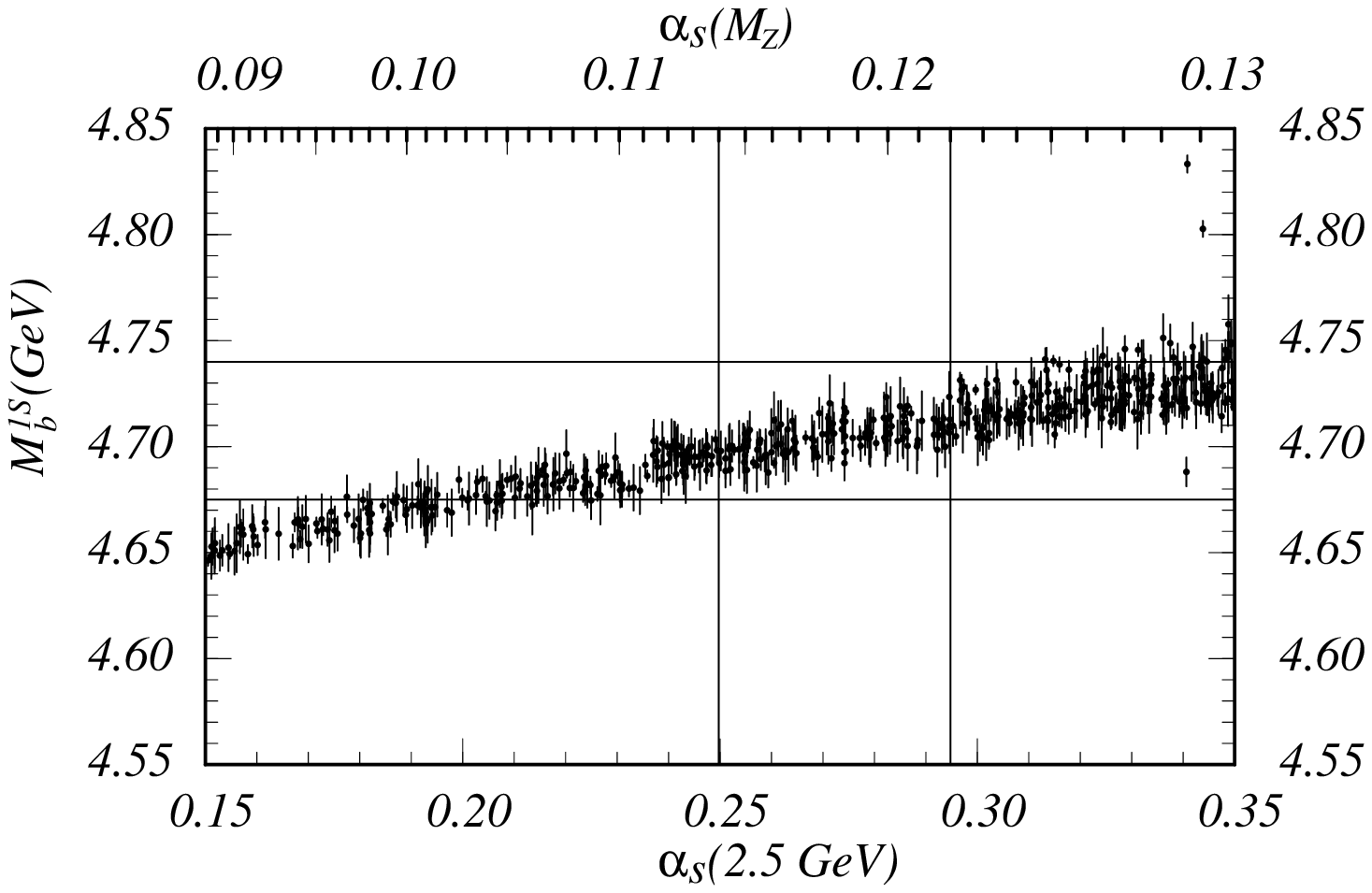}
%
\vskip  1.3cm
 \caption{\label{figsumrules}
The dark regions show the
allowed bottom mass values as a function of
$\alpha_s$ using NNLO $\Upsilon$ sum rules in the 
pole (upper figure), $\overline{\mbox{MS}}$ (middle) and $1S$ (lower
figure) mass schemes. 
The diagrams have
been generated from results obtained in~\cite{Hoang3,Hoang4}.
Mass extractions for $\alpha_s(M_Z)=0.118\pm 0.004$ are indicated. 
Sum rule analyses at NNLO have also been carried out
in~\cite{Melnikov1,Penin2,Beneke2}. 
}
 \end{center}
\end{figure}
In Figs.~\ref{figsumrules} the results for the 
pole (upper figure), $\overline{\mbox{MS}}$ (middle) and $1S$ (lower)
mass at NNLO from a 
simultaneous fit of four different moments\footnote{
In Ref.~\cite{Beneke2} this method has been criticised as being
incapable of estimating theoretical uncertainties, because it uses
moments for identical choices of the theoretical input parameters and 
because the form of the covariance matrix, which affects the
theoretical error, is determined from experimental data. This
criticism cannot be applied here because choosing input parameters for
the fitted moments independently is practically equivalent to only
fitting individual moments and because the way of estimating the
uncertainty by scanning the theoretical parameter space does in
general not allow for a clear separation of experimental and
theoretical uncertainties. It was also indicated in
Ref.~\cite{Beneke2} that the obtained central value for the mass
obtained from the fit would not be unique because the
$\chi^2$-function for simultaneous fits of several moments is not
linear in the quark mass. This criticism does not apply because the
result of the fit is unique as it searches for the minimal $\chi^2$
value. 
} 
with $4<n<10$ are displayed
as a function of the strong coupling. To obtain the error band many
individual 95\% CL fits have been carried out for random choices of
the other theoretical parameters. The width of the error bands is
dominated by variations of the renormalisation scale in the QCD
potential. The shown spread has been obtained by randomly choosing
values above $1.5$~GeV.
We see that the pole and the $\overline{\mbox{MS}}$ mass analysis lead 
to mass extractions which are strongly correlated to the value of
$\alpha_s$ and which have uncertainties of 100 MeV. In the
$1S$ scheme, on the other hand, the correlation to $\alpha_s$ and also
the uncertainty is much smaller. Using $\alpha_s(M_Z)=0.118\pm 0.004$
as input we obtain $M_b^{1S}=4.71\pm 0.03$~GeV for the $1S$ mass,
where the error should be considered as $1\sigma$. 

This result can be cross-checked by using the fact that twice the
$1S$ bottom mass is equal to the mass of the $\Upsilon(1S)$ meson up
to non-perturbative corrections:
\begin{equation}
M_b^{1S} \, = \, 
\mbox{$
\frac{1}{2} M_{\Upsilon(1S)} - 
\frac{1}{2}\Delta_{\Upsilon(1S)}^{\rm non-pert}
$}
\,.
\end{equation} 
Because quantitative calculations for 
$\Delta_{\Upsilon(1S)}^{\rm non-pert}$ do not exist yet one can only
estimate its size and treat the estimate as an uncertainty. Such
estimates, using e.g. the gluon condensate contribution to an
ultra-heavy quarkonium, indicate that
$\Delta_{\Upsilon(1S)}^{\rm non-pert}$ is not larger than 100
MeV (see e.g.~\cite{Hoang5}). This estimate leads to 
\begin{equation}
M_b^{1S} \, = \,
\mbox{$
4.73\pm 0.05\,\mbox{GeV}
$}
\,,
\label{upsilonestimate}
\end{equation} 
which is perfectly consistent with the much more complicated sum rule
determination. [(P)NRQCD counting rules indicate that non-perturbative
effects associated with retardation effects are of NNLO in the
nonrelativistic expansion in the $b\bar b$ system~\cite{NRQCDtalks};
i.e. they should be of the order of the $\epsilon^3$ terms shown in
Eq.~(\ref{msbarbottom}) which is consistent with
Eq.~(\ref{upsilonestimate}).] The error in Eq.~(\ref{upsilonestimate})
should be considered as $1\sigma$. In fact, we can also consider the
sum rule calculation, where non-perturbative effects are much smaller
than for the individual $\Upsilon(1S)$ bound state, as a confirmation that 
the non-perturbative contributions in the $\Upsilon(1S)$ mass are
indeed as small as mentioned before. We emphasise, however, that the
sum rule determination of $M_b^{1S}$ and the result obtained in
Eq.~(\ref{upsilonestimate}) are not independent. We use the result in
Eq.~(\ref{upsilonestimate}) for the rest of this talk.

\section{$1S$ MASS AND NON-$Q\bar Q$ SYSTEMS}

\subsection{The upsilon expansion}
\label{subsectionupsilon}

The series defining the $1S$ mass, Eq.~(\ref{M1Sdef}), starts with
order $\alpha_s^2$ because the binding energy of a Coulombic $Q\bar Q$
system is of order $M_Q v^2\sim M_Q \alpha_s^2$. This feature
raises the question, how the $1S$ mass has to be implemented into
calculations for non-Coulombic quantities. The guiding principle for
the implementation of the $1S$ mass into these systems is that the
cancellation of the most infrared sensitive contributions contained in
Eq.~(\ref{M1Sdef}) has to be guaranteed after elimination of the pole
mass. This is achieved by the upsilon expansion~\cite{Hoang5}. In the
upsilon 
expansion terms of order $\alpha_s^n$ in non-Coulombic quantities are
of order $\epsilon^n$, whereas in Eq.~(\ref{M1Sdef}) they are of order
$\epsilon^{n-1}$. To implement the $1S$ mass one then has to eliminate
the pole mass and expand in the parameter $\epsilon$, which is set to
one afterwards. In other words, the upsilon expansion combines those
orders where the maximal power of $n_l$ is
the same. Thus the upsilon expansion combines terms of different order
in $\alpha_s$. This unusual prescription can
be understood from the fact that the leading IR-sensitive
contributions in Eq.~(\ref{M1Sdef}) (i.e. those contributions
involving the highest power of $\beta_0$ in each order) contain powers  
of the logarithmic term $L=\ln(\mu/C_F\,a_s\,M_b^{pole})$. These
logarithmic terms exponentiate at larger orders, 
$\sum_{i=0}L^i/i!\approx \exp(L)=\mu/C_F\,a_s\,M_b^{pole}$, and
effectively cancel one power of $\alpha_s$~\cite{Hoang5}.
In the following I will present a number of examples showing that the
$1S$ scheme, using the upsilon expansion, leads to nicely converging
perturbative series. Keeping in mind that the $1S$ mass can be
determined very accurately, this makes the $1S$ mass a very useful
scheme for phenomenological applications.

\subsection{Inclusive B decays}

In Refs.~\cite{Hoang5} the $1S$ scheme has been applied for all inclusive B
decay rates and some exclusive ones. In this talk I only report on the
inclusive semileptonic $B\to X_c e\bar\nu$ and $B\to X_u e\bar\nu$
decay rates, which are relevant for the determination of the CKM
matrix elements $|V_{cb}|$ and $|V_{ub}|$. 

At order $\epsilon^2$ in the $1S$ scheme the inclusive semileptonic
$b\to u$ decay rate reads
\begin{eqnarray}
\lefteqn{
\Gamma_{B\to X_u e\bar\nu} \,=\, 
\mbox{$
\frac{G_F^2 |V_{ub}|^2}{192\pi^3}
  (M_b^{1S})^5 \times
$}
}
\label{buups}
\\  &&
\mbox{$
\times [ 1 - 0.115\epsilon 
  - 0.031 \epsilon^2 
- \frac{9\lambda_2 - \lambda_1}{2(M_b^{1S})^2} + \ldots ] 
$}
\,.
\nonumber
\end{eqnarray}
The order $\epsilon^2$ term is exact~\cite{Ritbergen1}.
For comparison, the first three terms in the brackets in the pole
mass scheme are 
$[1-0.17\epsilon-0.10\epsilon^2]$. 
Using the $\overline{\mbox{MS}}$ mass they read
$[1+0.30\epsilon+0.13\epsilon^2]$.
Using Eq.~(\ref{upsilonestimate}) for the $1S$ mass and
$\lambda_2 = 0.12\,{\rm GeV}^2$ and 
$\lambda_1 = (-0.25\pm0.25)\,{\rm GeV}^2$ for the chromomagnetic and
the kinetic energy matrix elements Eq.~(\ref{buups}) implies
\begin{eqnarray}
|V_{ub}| &=& 
\mbox{$
(3.04 \pm 0.08 \pm 0.08) \times 10^{-3} \times
$}
\nonumber\\
&& 
\mbox{$
\times \left( {{\cal B}(B\to X_u e\bar\nu)\over 0.001}
  {1.6\,{\rm ps}\over\tau_B} \right)^{1/2}
$}
\,.
\label{Vub}
\end{eqnarray}
The first error is obtained by assigning an
uncertainty in Eq.~(\ref{buups}) equal to the value of the $\epsilon^2$ term
and the second is from assuming the 50 MeV uncertainty in $M_b^{1S}$.
The scale dependence of $|V_{ub}|$ due to varying $\mu$
in the range $m_b/2< \mu <2m_b$ is less than 1\%.  The uncertainty in
$\lambda_1$ makes a negligible contribution to the total error.  It is
not easy to measure ${\cal B}(B\to X_u e\bar\nu)$ without significant
experimental cuts, for example, on the hadronic invariant
mass. Using the $1S$ scheme should reduce the uncertainties in such
analyses as well. 

At order $\epsilon^2$ the inclusive semileptonic $b\to c$ decay
rate reads~\cite{Hoang5}
\begin{eqnarray}
\lefteqn{
\mbox{$
\Gamma(B\to X_c e\bar\nu) = \frac{G_F^2 |V_{cb}|^2}{192\pi^3}
  (M_b^{1S})^5 \times 0.533 \times
$}
}
\label{bcups}
\\ &&
\mbox{$
[ 1 - 
  0.096\epsilon - 0.029_{\tiny \rm BLM}\epsilon^2 
- \frac{0.28\lambda_2 + 0.12\lambda_1}{\rm GeV^2} 
\ldots ].
$}
\nonumber
\end{eqnarray}
For comparison, the perturbation series in this relation, when written
in terms of the pole mass, is $[1 - 0.12\epsilon - 0.07_{\rm
  BLM}\epsilon^2]$. Eq.~(\ref{bcups}) implies
\begin{eqnarray}
|V_{cb}| &=& 
\mbox{$
(41.6 \pm 0.8 \pm 0.7 \pm 0.5) \times 10^{-3} \times
$}
\nonumber\\&& 
\mbox{$
\times \eta_{\rm QED} \left( {{\cal B}(B\to X_c e\bar\nu)\over0.105}\,
  {1.6\,{\rm ps}\over\tau_B}\right)^{1/2} 
$}
\,,
\label{Vcb}
\end{eqnarray}
where $\eta_{\rm QED}\sim1.007$ is the electromagnetic radiative
correction. The uncertainties come from assuming an error in
Eq.~(\ref{bcups}) equal to the $\epsilon^2$ term, the $0.25\,{\rm
  GeV}^2$ error in $\lambda_1$, and the 50 MeV error in $M_b^{1S}$,
respectively. The agreement of $|V_{cb}|$ with other determinations
(such as exclusive decays) provides an additional check that
nonperturbative corrections in $M_{\Upsilon(1S)}$ are indeed small.  

\subsection{The top width and $\Delta \rho$}

It is illustrative to also examine the QCD corrections to the top quark
width, $\Gamma_t$, and to the top quark corrections to $\Delta\rho$
parameter. One might think that using the $1S$ mass for the top decay
width might lead to 
an improvement similar to the inclusive B decay rates. However, one
has to keep in mind that the top decays into a real W boson, which
makes the rate only depending on the third power of the top mass, and
that the overall renormalisation scale is much higher, which makes the
strong coupling quite small. Thus, using the $1S$ scheme, we might
expect an improvement in the convergence of the series compared to the
pole mass scheme, but it is not clear whether the $1S$ mass beats the 
$\overline{\mbox{MS}}$ mass. (We also have to keep in mind that for
$\Gamma_t$ and $\Delta\rho$ even the pole mass leads to an acceptable 
behaviour of the perturbative series.)
For simplicity we only consider the limit
$M_W=0$. Choosing $\alpha_s=0.11$ at the scale of the top mass 
the top decay width in pole mass scheme
reads~\cite{Czarnecki1,Chetyrkin1}
$\Gamma_t(t\to bW)=\Gamma_0^{pole}[1-0.10\epsilon-0.02\epsilon^2]$,
where $\Gamma_0\equiv\frac{G_F m_t |V_{tb}|^2}{8\pi\sqrt{2}}$.
In the $1S$ mass scheme the series is
$\Gamma_t=\Gamma_0^{1S}[1-0.09\epsilon-0.01\epsilon^2]$,
and using the $\overline{\mbox{MS}}$ mass (at the
$\overline{\mbox{MS}}$ mass) we have 
$\Gamma(t\to bW)=\bar \Gamma_0[1-0.04\epsilon-0.003\epsilon^2]$.
We see that the $1S$ mass leads to a better convergence than
the pole scheme, but in the $\overline{\mbox{MS}}$ mass scheme we
arrive at the best result. 
The situation is similar for $\Delta\rho$. In the massless W limit
and for $\alpha_s=0.11$ at the top mass scale the QCD corrections to
the top mass contributions in the $\Delta\rho$ in the pole mass scheme
are $\Delta\rho=x_t^{pole}[1-0.098\epsilon-0.017\epsilon^2]$, where
$x_t\equiv 3\frac{G_F m_t^2}{8\sqrt{2}\pi^2}$.
In the $1S$ scheme we have
$\Delta\rho=x_t^{1S}[1-0.095\epsilon-0.014\epsilon^2]$, and using the
$\overline{\mbox{MS}}$ mass (at the $\overline{\mbox{MS}}$ mass)
the result reads
$\Delta\rho=\bar x_t[1-0.007\epsilon-0.007\epsilon^2]$. As for the
case of $\Gamma_t$ the $\epsilon^2$ term is a factor of two larger
in the $1S$ scheme than in the $\overline{\mbox{MS}}$ scheme. This
observation seems to favour the use of the 
$\overline{\mbox{MS}}$ mass, but, as we will show just below, 
it does not necessarily lead to smaller uncertainties because the
error in $\overline{\mbox{MS}}$ mass is always larger than the error
in the $1S$ mass.

\subsection{Determination of $\overline{\mbox{MS}}$ masses}

By construction, the $1S$ mass does not know much about large momenta
above the inverse Bohr radius $\sim M_Q\alpha_s$. Thus it cannot be
expected to serve as a practical mass prescription for high energy
processes where the renormalisation point is well above the
heavy quark mass. For those systems the $\overline{\mbox{MS}}$ mass is
undeniably one of the best choices. (A well known example is
$e^+e^-$ annihilation into massive quarks at high energies, where the
use of the $\overline{\mbox{MS}}$ mass leads to the cancellation of
certain large logarithms.) Nevertheless, an accurate
determination of the $1S$ mass can also provide a refined
determination of the $\overline{\mbox{MS}}$ mass at a low, but still  
reasonable, scale of order the heavy quark mass. The obtained result
for the $\overline{\mbox{MS}}$ mass can then be evolved up to the
characteristic scale of the high energy process.

To determine the $\overline{\mbox{MS}}$ mass from $M_b^{1S}$ we again
have to employ the upsilon expansion. We emphasise that in order to
obtain the relation between the $\overline{\mbox{MS}}$ mass and
$M_b^{1S}$ to order $\epsilon^3$ we need the relation between the
$\overline{\mbox{MS}}$ and the pole mass to three loops. Whereas the
one and two loop corrections are known for quite a while, a numerical
calculation of the three loop terms has just been announced on this
conference ~\cite{Chetyrkin2,Chetyrkin3}. The result obtained in 
Ref.~\cite{Chetyrkin3} reads
($\overline m\equiv \overline m^{(n_l+1)}$, 
$a\equiv\alpha_s^{n_l+1}(\overline m(\overline m))$)
\begin{eqnarray}
\lefteqn{
\mbox{$
M^{pole} \, = \, \overline m(\overline m)[\,
1 + 0.4243\, a \,\epsilon 
$}
}
\label{polemsbar}
\\&&
\mbox{$+ 
a^2\,(1.362 - 0.1055 n_l)\,\epsilon^2
$}
\nonumber
\\
\lefteqn{ 
\mbox{$\quad+
a^3\,( 6.26(16)  -  0.871(22) n_l +  0.02106 n_l^2  )\,\epsilon^3
\,]
\,.
$}
}
\nonumber
\end{eqnarray}
Combining Eqs.~(\ref{M1Sdef}) and (\ref{polemsbar}) we arrive at
($\bar a\equiv\alpha_s^{(n_l)}(M^{1S})$,  $\ell\equiv\ln\bar a$)
\begin{eqnarray}
\lefteqn{
\overline m(\overline m) \, = \, 
M^{1S}\{1 + 
[-0.4243 \bar a + 0.2222 \bar a^2] \epsilon
}
\nonumber
\\ 
\lefteqn{
+[
0.0494 \bar a^4 
+ \bar a^2 (-1.18 + 0.106 n_l) 
}
\nonumber
\\ 
\lefteqn{ \quad
+ \bar a^3 (0.825 - 
     0.778 \ell + (-0.0729 + 0.0472 \ell)n_l )
]\epsilon^2
}
\nonumber
\\ 
\lefteqn{
+[
\bar a^3 (-5.5(2) + 0.80(2) n_l - 0.021 n_l^2)
}
\nonumber
\\ 
\lefteqn{  \quad
+ \bar a^4 (5.8 - 3.7 \ell +  2.0 \ell^2
}
\nonumber
\\ 
\lefteqn{  \qquad + 
    (-0.69 + 0.56 \ell - 0.25 \ell^2 ) n_l
}
\nonumber
\\ 
\lefteqn{  \qquad +
    (0.020 - 0.018 \ell + 0.0075 \ell^2) n_l^2  )
}
\nonumber
\\ 
\lefteqn{  \quad
+ \bar a^5 (0.21 -  0.35 \ell + (-0.022 + 0.021 \ell)n_l ) 
}
\nonumber
\\ 
\lefteqn{  \quad
+ 0.011 \bar a^6 
] \epsilon^3
\}
\,.
}
\label{msbarm1S}
\end{eqnarray}

Using the value for bottom $1S$ mass displayed in
Eq.~(\ref{upsilonestimate}) we arrive at
\begin{eqnarray}
\lefteqn{
\overline m_b(\overline m_b) =
\mbox{$
[ 4.73 - 0.38 \,\epsilon - 0.10 \,\epsilon^2 -
0.04\, \epsilon^3
$}
}
\nonumber
\\& & \qquad
\mbox{$
 \pm 0.05 (\delta M_b^{1S})
 \pm x \,0.01 (\delta\alpha_s)]
$~GeV}
\,.
\label{msbarbottom}
\end{eqnarray} 
The good convergence of the terms in the upsilon expansion is expected
because both the $1S$ and $\overline{\mbox{MS}}$ masses are
short-distance masses, i.e. their relation does not have the bad high
order behaviour that plagues the pole mass. In Eq.~(\ref{msbarbottom})
also the error from the uncertainty in the strong coupling is
displayed assuming $\alpha_s(M_Z)=0.118\pm x\,0.001$. For $x=4$
this amounts to an error of 40~MeV, which is as large as the
$\epsilon^3$ term. This uncertainty is nothing else than the strong
correlation visible in the lower picture of Fig.~\ref{figsumrules}.
It cannot be eliminated, unless the uncertainty in the strong
coupling is reduced. What we gain, however, by using the $1S$ mass in
the first place, is a reduction of the uncertainties arising from the
variations of theoretical parameters like the renormalisation scale.
(Those uncertainties are associated with the width of the error band
in the lower picture of Figs.~\ref{figsumrules}.)
Combining the uncertainties quadratically
be arrive at 
\begin{eqnarray}
\overline m_b(\overline m_b) & = & 4.21 \, \pm \, 0.07~\mbox{GeV} 
\label{MSbarfinal}
\end{eqnarray} 
for the $\overline{\mbox{MS}}$ bottom quark mass. The uncertainty
should be considered as $1 \sigma$. For the $1S$ mass value obtained
from the sum rule analysis we get 
$\overline m_b(\overline m_b) = 4.19 \pm 0.06~\mbox{GeV}$. We
note that the central value can vary by up to 30 MeV depending on
which method one uses to determine the $\overline{\mbox{MS}}$ mass
from Eqs.~(\ref{M1Sdef}) and (\ref{polemsbar}). Thus a conversion
error of 30 MeV is included in the error estimates.

The same analysis could also be carried out for the top quark mass.
So let us assume that the LC has determined the top $1S$ mass 
as $M_t^{1S}=175\pm 0.2$~GeV from the threshold scan. The expression that
corresponds to Eq.~(\ref{msbarbottom}) then reads
$ \bar m_t(\bar m_t) = 
[ 175 - 7.50\, \epsilon - 
0.94\, \epsilon^2 - 
0.19\, \epsilon^3
\pm 0.2 (\delta M_{1S})
\pm x\,0.07 (\delta \alpha_s)
]$~GeV, 
which would imply 
$\bar m_t(\bar m_t) = 166.4 \pm 0.4$~GeV for $x=4$. 
Again, the uncertainty in the $\overline{\mbox{MS}}$ mass is 
larger unless the uncertainty in the strong coupling is significantly
reduced.

\section{OTHER MASS DEFINITIONS}

The idea of using a short-distance mass especially adapted to problems
with a low characteristic scale has not been new. In Ref.~\cite{Bigi1}
the so-called ``kinetic mass'', defined by a cutoff-dependent
subtraction from the heavy quark self energy, has been proposed and used to
parameterise inclusive B decays with results similar to the ones shown
in this talk. At present, the kinetic mass is only known to order
$\epsilon^2$~\cite{Czarnecki2}. An analysis using $\Upsilon$ sum rules 
determining the kinetic mass can be found in Ref.~\cite{Melnikov1}. In
Ref.~\cite{Beneke0} the so called ``potential-subtracted'' mass,
defined by a cutoff-dependent subtraction from the static QCD
potential, has been
proposed as a mass definition adequate for threshold problems. In
Ref.~\cite{Beneke1} the potential-subtracted mass has been employed to
describe $t\bar t$ production close to threshold at the LC and in
Ref.~\cite{Beneke2} it has been determined using $\Upsilon$ sum
rules.                     
However, because beyond order $\epsilon^3$ (=NNLO in the non-relativistic
expansion), the static potential is sensitive to scales below the
inverse Bohr radius~\cite{Appelquist1}, further considerations are 
required to fully justify its use as a short-distance mass. 

It should be emphasised that there are almost an infinite numbers of ways
to define low scale short-distance masses, so even more might be
invented in the future. 
In this respect it is very reasonable to use
the $\overline{\mbox{MS}}$ mass as a common reference point which
allows to check the consistency of those masses. (In fact, using
one of the low scale short-distance masses as a reference point would
be a smarter choice owing to the loss of precision in the
conversion to the 
$\overline{\mbox{MS}}$ mass scheme.)
In the case of bottom
quark mass determinations the obtained results for the
$\overline{\mbox{MS}}$ mass from $1S$ ($4.21\pm 0.07$~GeV), 
kinetic ($4.2\pm 0.1$~GeV~\cite{Melnikov1}) and
potential-subtracted mass ($4.25\pm 0.08$~GeV~\cite{Beneke2}) analyses
are perfectly consistent. A critical and comprehensive comparison to 
(and review of) all $\overline{\mbox{MS}}$ bottom mass
determinations prior to the ones just mentioned is certainly useful 
and shall be carried out elsewhere.  

\section{ACKNOWLEDGEMENTS}

I thank Z. Ligeti, A.~V. Manohar and T.~Teubner for
their collaboration on topics reported in this talk. 
This work is supported in part by the EU Fourth Framework Program
``Training and Mobility of Researchers'', Network ``Quantum
Chromodynamics and Deep Structure of Elementary Particles'', contract
FMRX-CT98-0194 (DG12-MIHT).

\end{document}